# Performance Evaluation of Mesh based Multicast Reactive Routing Protocol under Black Hole Attack

E.A.Mary Anita
Research Scholar
Anna University
Chennai

V.Vasudevan
Senior Professor and Head / IT
A. K. College of Engineering
Virudunagar, India

*Abstract*— A mobile ad-hoc network is an autonomous system of mobile nodes connected by wireless links in which nodes cooperate by forwarding packets for each other thereby enabling communication beyond direct wireless transmission range. The wireless and dynamic nature of ad-hoc networks makes them vulnerable to attacks especially in routing protocols. Providing security in mobile ad-hoc networks has been a major issue over the recent years. One of the prominent mesh base reactive multicast routing protocols used in ad-hoc networks is On Demand Multicast Routing protocol (ODMRP). The security of ODMRP is compromised by a primary routing attack called black hole attack. In this attack a malicious node advertises itself as having the shortest path to the node whose packets it wants to intercept. This paper discusses the impact of black hole attack on ODMRP under various scenarios. The performance is evaluated using metrics such as packet delivery ratio and end to end delay for various numbers of senders and receivers via simulation. Simulations are carried out using network simulator ns-2. The results enable us to propose solutions to counter the effect of black hole attack.

## I. INTRODUCTION

Security in wireless ad-hoc networks is a complex issue. This complexity is due to various factors like insecure wireless communication links, absence of a fixed infrastructure, node mobility and resource constraints [1]. Nodes are more vulnerable to security attacks in mobile ad-hoc networks than in traditional networks with a fixed infrastructure. The security issues of Mobile Ad-hoc Networks (MANETs) are more challenging in a multicasting environment with multiple senders and receivers. There are different kinds of attacks by malicious nodes that can harm a network and make it unreliable for communication. These attacks can be classified as active and passive attacks [2]. A passive attack is one in which the information is snooped by an intruder without disrupting the network activity. An active attack disrupts the normal operation of a network by modifying the packets in the network. Active attacks can be further classified as internal and external attacks. External attacks are carried out by nodes that do not form part of the network. Internal attacks are from compromised nodes that were once legitimate part of the network.

A black hole attack is one in which a malicious node advertises itself as having the shortest path to a destination in a network. This can cause Denial of Service (DoS) by dropping the received packets.

The rest of the paper is organized as follows. The next section gives an overview of ODMRP. Section III discusses about black hole attack. Section IV over views security in ad-hoc networks. In section V the results of simulation experiments that show the impact of black hole attack on the performance of ODMRP under scenarios are discussed. Finally section VI summarizes the conclusion.

## II. OVERVIEW OF ODMRP

ODMRP is a mesh based multicast routing protocol that uses the concept of forwarding group. Only a subset of nodes forwards the multicast packets on shortest paths between member pairs to build a forwarding mesh for each multicast group [3].

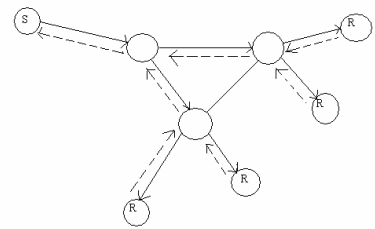

O – Mobile node
S – Multicast Source
R – Multicast Receiver
______ JREQ
_ _ _ _ _ JREP

Figure 1 On demand route and mesh creation

In ODMRP, group membership and multicast routes are established and updated by the source on demand. When a multicast source has packets to send, it initiates a route discovery process. A JOIN REQUEST packet is periodically broadcast to the entire network. Any intermediate node that receives a non- duplicate JREQ packet stores the upstream node ID and rebroadcasts the packet. Finally when this packet reaches the destination, the receiver creates a JOIN REPLY and broadcasts it to its neighbors. Every node receiving the JREP checks to see if the next node id in JREP matches its own. If there is a match, it is a part of the forwarding group, sets its FG_FLAG and broadcasts its JREP built upon matched entries. This JREP is thus propagated by each forwarding group member until it reaches the source via a shortest path. Thus routes from sources to receivers build a mesh of nodes called forwarding group.

The forwarding group is a set of nodes that forward the multicast packets. It supports shortest paths between any member pairs. All nodes inside the bubble (multicast members and forwarding group nodes) forward multicast data packets. A multicast receiver can also be a forwarding group node if it is on the path between a multicast source and another receiver. The mesh provides richer connectivity among multicast members compared to trees.

After the route establishment and route construction process, a multicast source can transmit packets to receivers via selected routes and forwarding groups. A data packet is forwarded by a node only if it is not a duplicate one and the setting of the FG_Flag for the multicast group has not expired. This procedure minimizes traffic overhead and prevents sending packets through stale routes.

In ODMRP, no explicit control packets need to be sent to join or leave the group. A multicast source can leave the group by just stop sending JREQ packets when it does not have any data to be sent to the group. If a receiver no longer wants to receive data from a particular group, it removes the corresponding entries from its member table and does not transmit the JOINTABLE for that group.

## III. BLACK HOLE ATTACK

A black hole attack is one in which a malicious node uses the routing protocol to advertise itself as having the shortest path to the node whose packets it wants to intercept[4]. This attack aims at modifying the routing protocol so that traffic flows through a specific node controlled by the attacker. The attacker drops the received messages instead of relaying them as the protocol requires. Therefore the quantity of routing information available to other nodes is reduced. The attack can be accomplished either selectively or in bulk. Selective dropping means dropping packets for a specified destination or a packet every 't' seconds or a packet every 'n' packets or a randomly selected portion of packets[5]. Bulk attack results in dropping all packets. Both result in degradation in the performance of the network.

### A. Black hole problem in ODMRP

ODMRP is an important on demand routing protocol that creates routes only when desired by the source node. ODMRP does not include any provisions for security and hence it is susceptible to attacks. When a node requires a route to a destination it initiates a route discovery process within the network. Any malicious node can interrupt this route discovery process by claiming to have the shortest route to the destination thereby attracting more traffic towards it. For example, source A wants to send packets to destination D, in fig.1, source A initiates the route discovery process. Let M be the malicious node which has no fresh route to destination D. M claims to have the route to destination and sends join reply JREP packet to S. The reply from the malicious node reaches the source node earlier than the reply from the legitimate node, as the malicious node does not have to check its routing table like the other legitimate nodes. The source chooses the path provided by the malicious node and the data packets are dropped. The malicious node forms a black hole in the network and this problem is called black hole problem. called black hole problem.

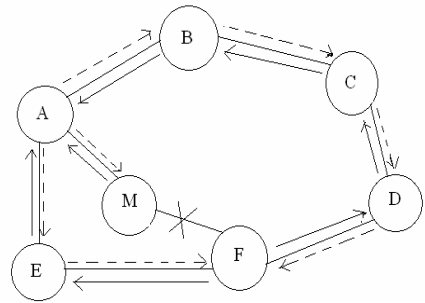

A-Source node
D-Destination node
M-Malicious node
- - - - JREQ
____ JREP

Figure 2 Black hole attack

## IV. RELATED WORK

Several researchers have addressed the problem of securing unicast routing protocols for ad-hoc networks. Ariadne [7], Secure Aware Ad-Hoc Routing (SAR) [8], Secure Efficient Ad-Hoc Distance (SEAD) Vector Routing Protocol [9], Secure AODV [10], Authenticated Routing for Ad-Hoc Network (ARAN) [11], Secure Routing Protocol (SRP) [12] and Secure Link- State Protocol (SLSP) [12] are all based on unicast routing protocols. Also these protocols do not address the problem of black hole attack. Marti, S., Giuli, T. J., Lai, K., & Baker, M.[13] have proposed a Watchdog and Pathrater approach against black hole attack which is implemented on top of source routing protocol such as DSR (Dynamic Source Routing). Ramanujam et al. [2] have presented some general techniques collectively called as TIARA (Techniques for Intrusion resistant Ad-Hoc Routing Algorithms) to protect ad-hoc networks from attacks.

CONFIDANT (Cooperative of Nodes, Fairness In Dynamic Ad-hoc Networks) [10] is an extended version of Watchdog and Pathrater which uses a mechanism similar to Pretty Good Privacy for expressing various levels of trust, key validation and certification. It is also implemented on unicast routing protocol such as DSR. These papers have not addressed the challenges in multicast routing protocols which are our focus in this paper.

## V. PERFORMANCE EVALUATION

The performance of a network depends on many factors such as number of senders, receivers, attackers and their positions. The performance of ODMRP has been observed in different scenarios.

### A. Simulation Environment and Metrics

The simulation is done using the ns-2 simulator. The metrics used in evaluating the performance are:
Packet Delivery Ratio: The ratio of the number of data packets delivered to the destinations to the number of data packets generated by the sources.
Average End-to-End Delay: This is the average delay between the sending of packets by the source and its receipt by the receiver. This includes all possible delays caused during data acquisition, route discovery, queuing, processing at intermediate nodes, retransmission delays, propagation time, etc. [5]. It is measured in milliseconds.

### B. Simulation Profile

The simulation settings are as follows. The network consists of 50 nodes placed randomly within an area of 1000m x 1000 m. Each node moves randomly and has a transmission range of 250m. The random way point model is used as the mobility model. In this model, a node selects a random destination and moves towards that destination at a speed between the pre-defined maximum and minimum speed. The minimum speed for the simulations is 0 m/s while the maximum speed is 50 m/s. The simulations were carried out with 2, 5, 7 and 9 attackers for different number of receivers. The malicious nodes were selected randomly.

### C. Discussion of results

Fig.1 shows the variation of PDR with mobility for 1 sender and 20 receivers when the number of attackers are varied from 0 to 5. It is seen that the PDR decreases with increased mobility.

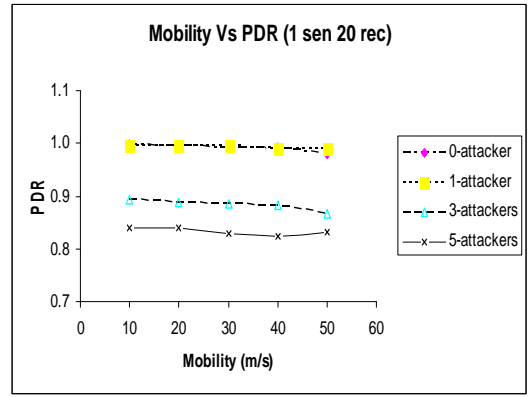

Figure 3 PDR for 1 sender and 20 receivers

The drop in PDR in the presence of a single attacker is only around 1%. When the number of attackers is increased to 3, the drop in PDR increases by 10% and a further drop of 5% is observed when the number of attackers is increased to 5.Higher the number of attackers, higher the reduction in PDR.

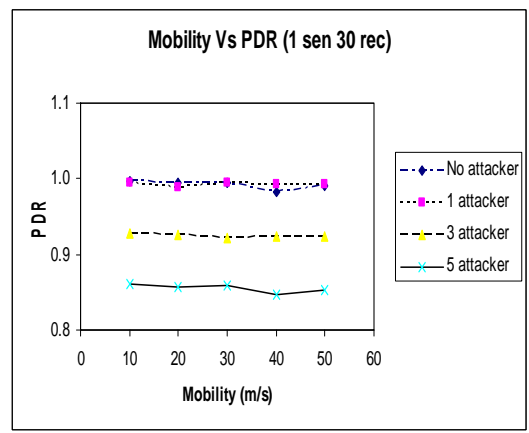

Figure 4 PDR for 1 sender and 30 receivers

A similar situation in seen in fig. 4 also but, as the number of receivers in this case is increased to 30, the impact of the attack is comparatively lesser. This is due to the fact that more number of receivers results in a denser routing mesh providing alternate paths for the data packets. Given the same number of attackers, the PDR is higher for higher number of receivers.

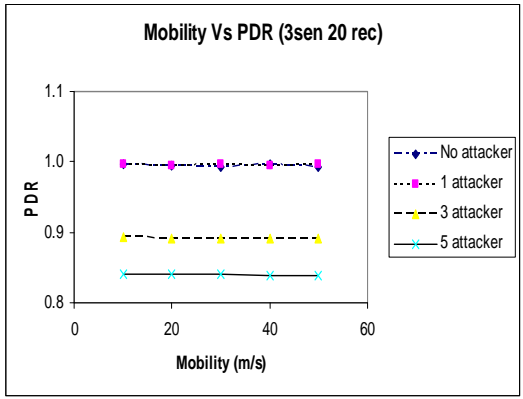

Figure 5 PDR for 3 senders and 20 receivers

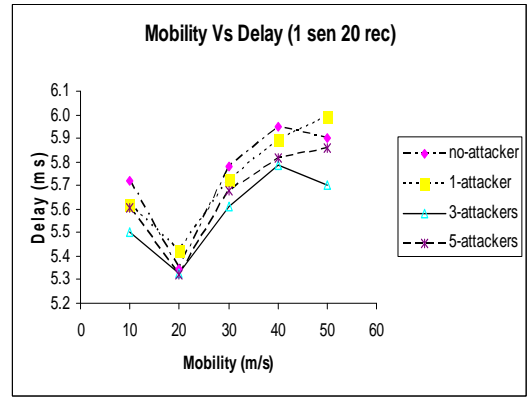

Figure 7 Delay for 1 sender and 20 receivers

Fig. 5 shows an increase in the value of PDR compared to fig. 3. This can be attributed to the increased number of senders thereby providing more alternate paths for the data packets. Even if a packet gets dropped in one path due to the presence of black hole nodes, there is a chance for the duplicate copy of the packet to reach the destination through alternate paths free from malicious nodes.

Fig.7 shows the variation of end to end delay for different number of attackers in the presence of 1 sender and 20 receivers. There seems to be an increase in the delay in the presence of attackers.

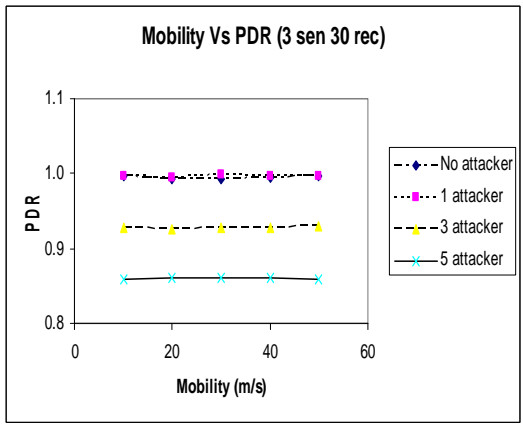

Figure 6 PDR for 3 senders and 30 receivers

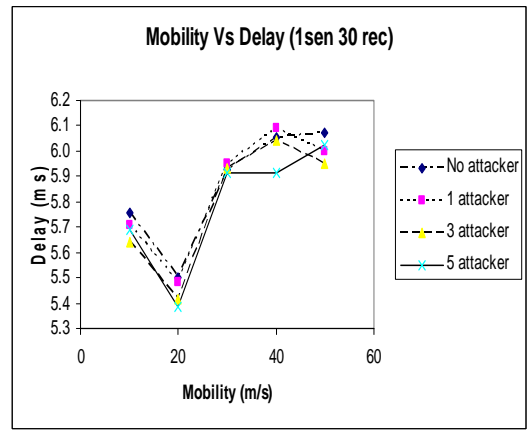

Figure 8 Delay for 1 sender and 30 receivers

The increase in PDR when the number of receivers is increased from 20 to 30 with the same number of senders varies from 1% in the absence of attackers to 3% in the presence of 5 attackers. This is clearly depicted in fig.6.

From the above graphs we conclude that a large multicast group with more number of senders and receivers are more resilient to black hole attack than a smaller group. This is due to the presence of more alternative paths available to route duplicate copies of the data packets.

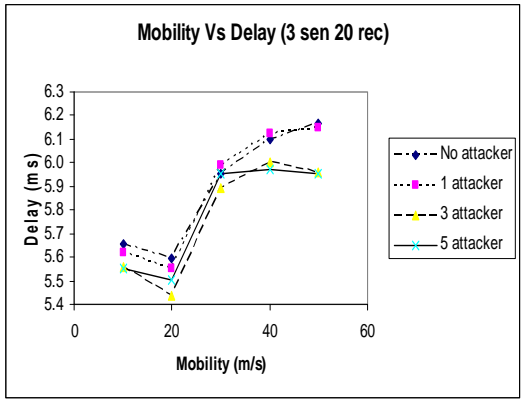

Figure 9 Delay for 3 senders and 20 receivers

This is due to the fact that non shortest paths containing black hole nodes are selected for routing the packets. Also we see that the delay increases with increased group size as shown in fig. 8.

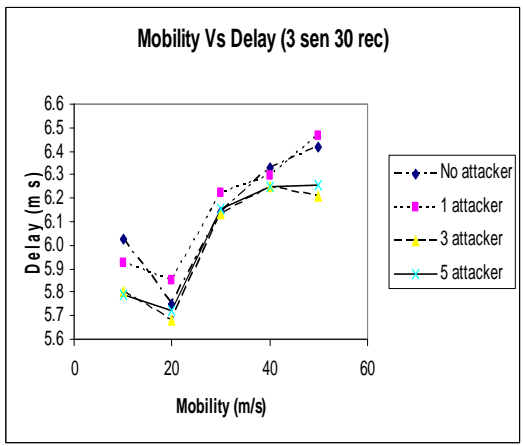

Figure 10. Delay for 3 senders and 30 receivers

End to end delay includes all delays caused during route discovery, transmission delays, processing at intermediate nodes, etc. Obviously, a larger group accounts to a larger delay. This is clearly depicted in fig. 9 and fig. 10.

## VI. CONCLUSION

Security is one of the major issues in MANETs. In this paper the effect of black hole attack on MANETs has been analysed. The multicast routing protocol ODMRP has been simulated with black hole nodes under different scenarios. The performance of a multicast routing protocol under black hole attack depends on factors such as number of multicast senders, number of multicast receivers and number of black hole nodes

From the simulation results it is observed that, the packet delivery ratio reduces with increased mobility of the nodes and also with increased number of black hole nodes and affect the performance of the network. Also the packet delivery ratio is higher for large number of receivers for the same number of attackers. That is, the effect of the attack is more in a small group than in a large group. A large group is able to withstand the attack to a reasonable extent when compared to a smaller group which is easily susceptible to attacks. This can be attributed to the existence of alternate paths for routing the data packets.

The results also depict that the delay increases with increase in group size and increase in number of attackers. This is because of the fact that non shortest paths containing black hole nodes are selected for routing the packets.

To implement security over ODMRP, all route request messages are to be authenticated. Several mechanisms can be found in literature for authentication. A self organized public key infrastructure can be used to authenticate the nodes participating in the route discovery process so that compromised nodes can be easily identified and excluded from the network.

## REFERENCES

[1] D. Djenouri, L. Khelladi, and N. Badache, "A Survey of Security Issues in Mobile Ad Hoc and Sensor Networks," IEEE Communication .Surveys & Tutorials, vol. 7, no. 4, 4th Quarter 2005

[2] L. Zhou and Z. J. Haas, "Securing Ad Hoc Networks," IEEE Network Magazine., vol. 13, no.6, Nov./Dec. 1999, pp. 24–30.

[3] S.Lee, M.Gerla and C.Chain, "On Demand Multicast Routing protocol-(ODMRP)," Proc. of the IEEE Wireless Communication and Networking Conference (WCNC), September 1999.

[4] P. Papadimitratos and Z. J. Haas, "Secure Routing for Mobile Ad hoc Networks," Proc. Communication Networks and Distributed Systems, Modeling and Simulation Conf. (CNDS'02), San Antonio, Texas, Jan. 2002, pp. 27–31.

[5] P. Ning and K. Sun, "How to Misuse AODV: A Case Study of Insider Attacks Against Mobile Ad Hoc Routing Protocols," Info. Assurance Workshop, IEEE Sys., Man and Cybernetics Soc., June 2003, pp. 60–67.

[6] H. Deng, W. Li, and Dharma P. Agrawal, "Routing security in Ad Hoc Networks," IEEE Communications Magazine, Special Topics on Security in Telecommunication Networks, Vol. 40, No. 10, October 2002, pp. 70-75.

[7] Y.-C. Hu, A. Perrig, and D. Johnson. Ariadne: "A secure on-demand routing protocol for ad hoc networks," Proc. of 8th ACM Mobile Computing and Networking (MobiCom'02), pp. 12–23, 2002.

[8] S. Yi, P. Naldurg, and R. Kravets, " Security-aware ad hoc routing for wireless networks,". Proc. of 2nd ACM

Mobile Ad Hoc Networking and Computing (MobiHoc'01), pp. 299–302, 2001.
[9] Y.-C. Hu, D. Johnson, and A. Perrig. SEAD: "Secure efficient distance vector routing in mobile wireless ad hoc networks." Proc. of 4$^{th}$ IEEE Workshop on Mobile Computing Systems and Applications (WMCSA'02), pp. 3–13, 2002
[10] Yang, H., Luo, H., Ye, F., Lu, S., & Zhang, L. (2004). "Security in mobile ad hoc networks: Challenges and solutions," IEEE Wireless Communications, 11(1), 38-47.
[11] Sanzgiri, K., Dahill, B., Levine, B. N., Shields, C., & Belding-Royer, E. M. (2005). "Authenticated routing for ad hoc networks," IEEE Journals on Selected Areas in Communications, 23(3), 598- 610
[12] Papadimitratos, P., & Haas, Z. J. (2003a). "Secure link state routing for mobile ad hoc networks." Proceedings of the Symposium on Applications and the Internet Workshops (SAINT) (pp. 27-31).
[13] Marti, S., Giuli, T. J., Lai, K., & Baker, M. (2000). "Mitigating routing misbehavior in mobile ad-hoc networks," Proceedings of the 6th International Conference on Mobile Computing and Networking (MobiCom), ISBN 1-58113-197-6 (pp. 255-265).

AUTHORS PROFILE

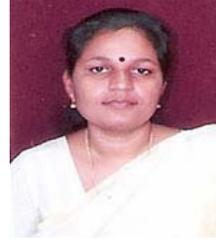

**E.A.Mary Anita** received the Bachelor of Engineering in Electrical and Electronics Engineering and Master of Engineering in Computer Science and Engineering, from Government College of Engineering, Tirunelveli, TamilNadu, affiliated to Madurai Kamaraj University and Manonmaniam Sundaranar University respectively. Her research interests include wireless communication, multicast and network security.

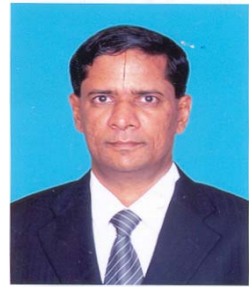

**V.Vasudevan** received the PhD degree from Madurai Kamaraj University, India. He is Senior Professor in the Information Technology Department of A.K.College of Engineering, Virudhunagar, TamilNadu. His research interests include multicasting, image processing and grid computing